**Stabilizing γ-$MgH_2$ at Nanotwins in Mechanically Constrained Nanoparticles**

Jochen A. Kammerer, Xiaoyang Duan*, Frank Neubrech, Rasmus R. Schröder, Na Liu*, Martin Pfannmöller*

J.A.K. Author 1, 3DMM2O, Cluster of Excellence (EXC-2082/1 – 390761711) and CAM – Centre for Advanced Materials, Heidelberg University, Im Neuenheimer Feld 225, 69120 Heidelberg, Germany

Dr. X.D. Author 2, Dr. F.N. Author 3, Prof. N.L. Author 5, MPI – Max Planck Institute for Solid State Research, 70569 Stuttgart, Germany
E-mail: duan@is.mpg.de

Dr. F.N. Author 3, Prof. N.L. Author 5, 2nd Physics Institute, University of Stuttgart, Pfaffenwaldring 57, 70569 Stuttgart, Germany
E-Mail: na.liu@pi2.uni-stuttgart.de

Prof. R.R.S. Author 4, 3DMM2O, Cluster of Excellence (EXC-2082/1 – 390761711) and Cryo Electron Microscopy, BioQuant, University Heidelberg, University Hospital, Im Neuenheimer Feld 267, 69120 Heidelberg, Germany

Dr. M.P. Author 6, CAM – Centre for Advanced Materials, Heidelberg University, Im Neuenheimer Feld 225, 69120 Heidelberg, Germany
E-mail: m.pfannmoeller@fastmail.net

**Abstract**

Reversible hydrogen uptake and the metal/dielectric transition make the system $Mg/MgH_2$ a prime candidate for solid state hydrogen storage and dynamic plasmonics. However, high dehydrogenation temperatures and slow dehydrogenation hamper broad applicability. One promising strategy to improve dehydrogenation is the formation of metastable $\gamma$-$MgH_2$. We present a nanoparticle (NP) design, where $\gamma$-$MgH_2$ forms intrinsically during hydrogenation and propose a formation mechanism based on transmission electron microscopy results. Volume expansion during hydrogenation causes compressive stress within the confined, anisotropic NPs, leading to plastic deformation of $\beta$-$MgH_2$ via $(301)_\beta$ twinning. These twins proposedly nucleate $\gamma$-$MgH_2$ nanolamellas, which are stabilized by residual compressive stress. Understanding this mechanism is a crucial step towards cycle stable, Mg based dynamic plasmonic and hydrogen storage materials with improved dehydrogenation. We envision that a more general design of confined NPs utilizes the inherent volume expansion to reform $\gamma$-$MgH_2$ during each rehydrogenation.

## 1. Introduction

Hydrogen is on the edge to become a major and carbon neutral energy carrier. However, handling the lightest element in the liquid or gaseous state is intricate, as it requires low temperature or high pressure. These central points are overcome by solid state storage materials. Taking capacity, reversibility, toxicity and abundancy into account, $Mg/MgH_2$ and its alloys are a solid choice for hydrogen storage, stimulating vibrant research activity.[1–4] Yet, for broad applicability, dehydrogenation has to be improved: $\beta$-$MgH_2$, the thermodynamically stable configuration of $MgH_2$, and Bulk Mg are in equilibrium at about 282 °C and 1 bar hydrogen pressure[5], requiring high operation temperatures for storage devices. Strategies to optimize dehydrogenation temperature, pressure and speed by thermodynamical destabilization of $MgH_2$ and lowered kinetic barriers include catalysts and alloying[6], nanostructuring[2,7,8], advanced nanocomposites[9–11], and the stabilization of metastable phases. However, many promising materials and structures are unstable and degrade during cycling due to high process temperatures and large volumetric changes (31.4 % Mg to β-$MgH_2$) during (de)hydrogenation.[6] Notably, Mg nanoparticles (NPs) have met interest for dynamic plasmonic applications[12–15] with similar requirements during cycling: fast (de)hydrogenation and conserved structural integrity. Especially stabilizing the metastable high pressure γ phase is a promising approach to lower the dehydrogenation temperature[16–18] but sustained generation of γ-$MgH_2$ during cycling has not been achieved.

Previously, insights into the dehydrogenation behaviour of $MgH_2$ materials were provided by *in-situ* transmission electron microscopy (TEM). These investigations comprise thin films[19], nanofibers[20], and nanopowders from ball milling[21–25]. We use TEM to decipher structural mechanisms in magnesium NPs from electron beam lithography (EBL) with reproducible size, shape, and defect structures. This allows to consistently compare the pristine, hydrogenated and dehydrogenated states. We followed the dehydrogenation *in-situ* by electron spectroscopic imaging (ESI)[21,23] and identified relevant structural features. We substantiated these findings by nano beam electron diffraction (NBD)[26] and electron tomography. We show: (i) Void formation by Mg diffusion is the dominant mechanism of structural degradation of mechanically constraint NPs during dehydrogenation. (ii) Excessive twinning leads to nanostructuring of the hydrogenated NPs. (iii) Nanolamellas of metastable γ-$MgH_2$ are stabilized within the twinned areas. Both, twinning and γ-$MgH_2$ improve dehydrogenation. They jointly form during hydrogenation as a result of volume expansion, mechanic constraints from a MgO belt, and anisotropy of the NPs, leading to high compressive stress and plastic deformation via twinning. The twins proposedly nucleate the high pressure γ-phase. Understanding these mechanisms will be crucial in advancing Mg-NP designs for dynamic plasmonics and hydrogen storage, as they inherit the potential to reform nanotwins and metastable γ-$MgH_2$ during each rehydrogenation.

## 2. Results

### 2.1. Characterization of the Nanoparticles

We investigated Mg NPs with a thickness of about 50 nm and diameters from 100-400 nm prepared from electron beam lithography and electron beam evaporation (see **Figure 1a**). The Ti buffer layers at the top and bottom of the pristine NPs mechanically decouple the Mg layer from the Pd capping layer and the amorphous $Si_3N_4$ substrate, leading to quasi-free NPs.[8,27] The catalytic Pd layer enables hydrogenation at ambient pressure.[28] As the Pd and Ti layers only cover the top and bottom of the NPs, a belt of native MgO forms around their edge.[29] Typical grain sizes of the Mg layer range from 40-90 nm (see **Figure 1b**). In agreement with previous findings[14], the NPs are textured with the $[0001]_{Mg}$ direction of the Mg grains normal to the substrate. This is evident from the diffraction pattern in **Figure 1c**. All strong reflections belong to the $[0001]_{Mg}$ zone axis ($10\text{-}10_{Mg}$, $11\text{-}20_{Mg}$ and $20\text{-}20_{Mg}$), whereas all other reflections are faint ($0002_{Mg}$, $10\text{-}11_{Mg}$, $10\text{-}13_{Mg}$ and $11\text{-}22_{Mg}$) or absent ($10\text{-}12_{Mg}$, $20\text{-}21_{Mg}$ and $0004_{Mg}$).

In the hydrogenated state, the diffraction contrast of the multiple Mg grains is lost and the structure of the NPs appears more homogeneous (compare pristine in **Figure 1b**, and hydrogenated **Figure 3** top left and section 11 in the supporting information (SI-11)). This implies the formation of $MgH_2$ from few nuclei only, which grow and consume the Mg phase. Only parallel lamellar contrasts appear as visible inhomogeneities within the hydrogenated NPs.

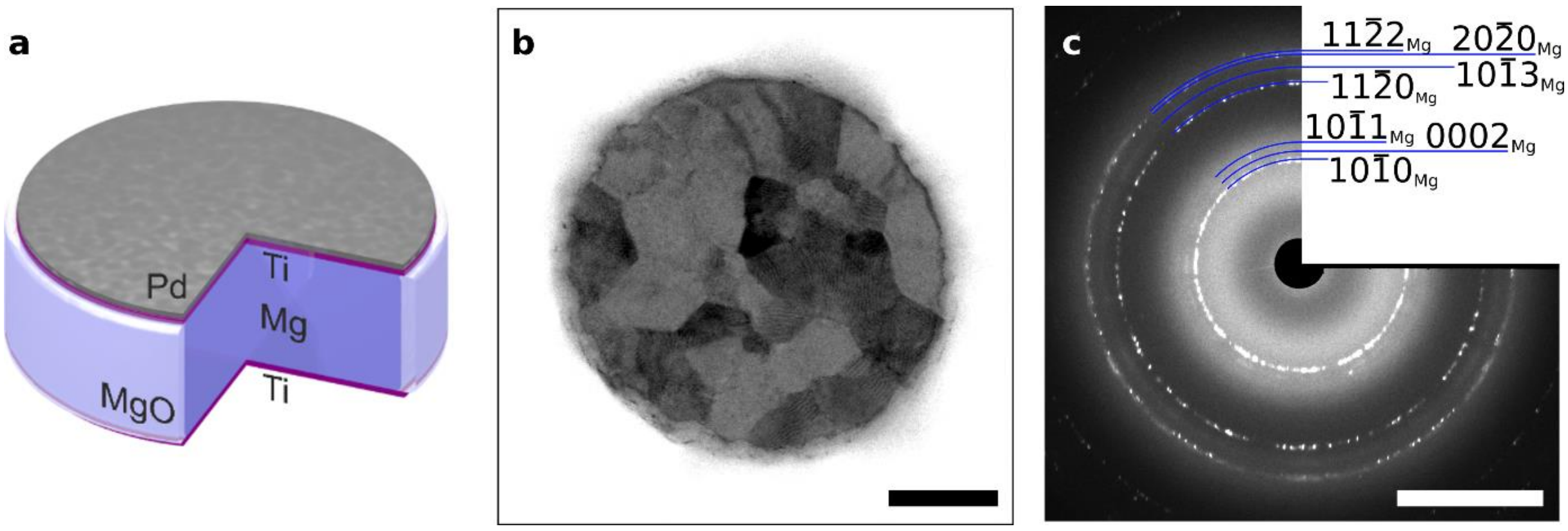

**Figure 1.** Mg nanoparticles in the pristine state. **a**, Schematic of a pristine nanoparticle from electron beam lithography. **b**, Zero loss filtered BF TEM image of a pristine Mg nanoparticle @200 kV; scale: 100 nm. **c**, Electron diffraction pattern of several pristine Mg nanoparticles. The absent of the expected MgO reflections are discussed in SI-2; scale: 5 $nm^{-1}$.

### 2.2. In-Situ Dehydrogenation

$MgH_2$ is beam sensitive and radiolytically decomposes in the TEM.[30] This allows to dehydrogenate $MgH_2$ NPs *in-situ* (see experimental design in **Figure 2**). In the hydrogenated state, the electron energy loss spectrum (EELS) is dominated by the broad $MgH_2$ volume plasmon peak at about 15 eV (**Figure 2b**).[23] This peak diminishes due to the decomposition of $MgH_2$ to Mg and $H_2$ with increasing exposure time. Simultaneously, the sharp Mg volume plasmon peak arises at about 10.6 eV, together with the minor and broad Mg surface plasmon peak around 8 eV energy loss. Said EELS signals can be exploited to spatially map the chemical composition by ESI. This allowed us to follow the *in-situ* dehydrogenation of $MgH_2$ NPs spatially and time resolved.

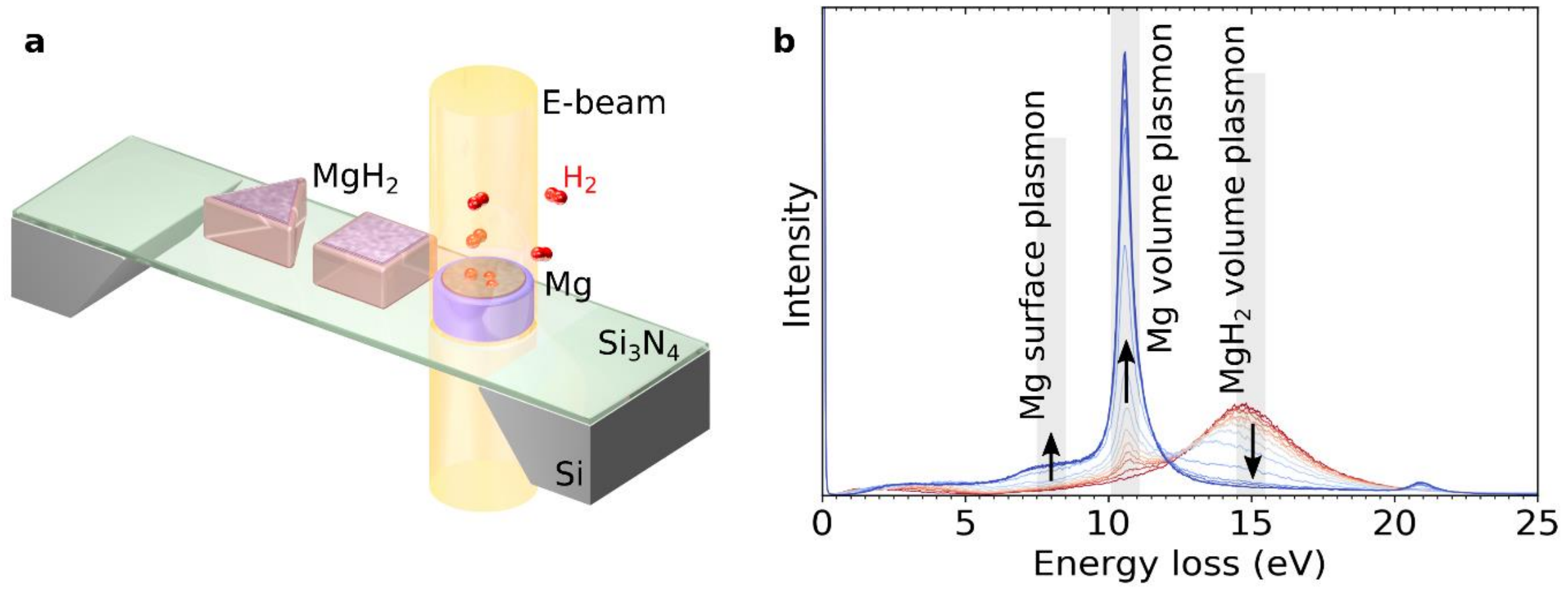

**Figure 2.** Experimental design. **a**, *In-situ* dehydrogenation of $MgH_2$ nanoparticles by radiolysis in the TEM. **e**, Background removed EEL spectra of a NP going from the hydrogenated (red) to the dehydrogenated state (blue). The three mayor plasmonic excitations are indicated together with their change of intensity during dehydrogenation (arrows). The bars indicate the energy windows for ESI with a slit width of 1 eV (see Figure 3).

An ESI dehydrogenation time series of a 350 nm NP is shown in **Figure 3**. We used an energy selective slit of 1 eV width to acquire zero loss filtered bright field images, and to visualize the excitation of the Mg volume, the Mg surface and the $MgH_2$ volume plasmons (see **Figure 2b**, bars**)**. We repeated this sequence continuously to dehydrogenate the NPs radiolytically and simultaneously monitor the related changes in structure and composition. Note that the images show signal intensities, which do not directly translate to the corresponding component. For example, the $Si_3N_4$ volume plasmon signal yields an intense background when imaging magnesium hydride. Furthermore, the objective aperture introduced diffraction contrast, which is superimposed with the EEL signal. Refer to SI-3 and SI-8 for further discussions. Nevertheless, as shown, the signal maps provide reliable, time-resolved chemical information about the NPs.

The MgO belt is visible in the bright-field images (**Figure 2**) as a brighter ring surrounding the NP. It appears as dark ring surrounded by bright halos in all ESI images. This agrees with the strong surface and weak volume plasmon excitation within the selected energy windows (see SI-4). The MgO belt is not altered during dehydrogenation.

When comparing the NP in the hydrogenated and dehydrogenated state (**Figure 3** left and right column), the $MgH_2$ volume plasmon signal has vanished and been replaced by an intense Mg volume plasmon signal. This Mg signal is not uniform throughout the NP but shows areas of high, intermediate and zero intensity. The latter two coincide with areas of strong Mg surface signals. The absence of Mg and the presence of Mg surface within the particle are caused by a network of voids, which forms during dehydrogenation. The voids penetrate the NP completely in normal direction, where the Mg signal is absent. Even though the NPs are considered quasifree[14,27], they are still confined by their MgO belt, which impedes free expansion and contraction. Thus, small cracks form to accommodate the volume shrinkage as soon as small amounts of the less voluminous Mg phase is generated (**Figure 3**, 2$^{nd}$ row). The evolution of voids can be directly followed by the excitation of the Mg surface plasmon (**Figure 3**, 3$^{rd}$ row). The excitation of the Mg surface plasmon is strong at the newly formed clean Mg-vacuum interface. Yet, it is aloof and delocalized over a considerable distance from the interface.[31] Therefore, and since surface plasmons cannot be excited within the Mg due to the *begrenzungs* effect, the Mg surface plasmon signal allows for a precise location of voids.[32] Refer to SI-12 for more details on the anticorrelation between the Mg surface and volume plasmon signals.

The cracks quickly compact, facet and grow by Mg diffusion to reduce surface energy. This ripening during ongoing dehydrogenation leads to the fine network of voids after about 3 min and the coarser network after about 6 min beam exposure. Similar voids form without beam exposure after the dehydrogenation of the NPs in air[12], excluding beam damage as underlying mechanism (see SI-6). The 3D-reconstruction of a dehydrogenated NP from electron tomography (**Figure 4**) shows that voids form predominantly at the edge and bottom. Void form is size independent within the investigated diameter range from 100-400 nm (see SI-7). Dehydrogenation is initially fastest at grain boundaries, which we identified from different diffraction contrast between the adjacent grains in the BF images (**Figure 3**, arrow). In terms of dehydrogenation speed, the grain boundaries are closely followed by particle edges, especially where they merge with the lamellar structures (**Figure 3**, dashed lines). Examining the volume of the NPs, dehydrogenation is generally faster within the nanolamellar structured parts: The $MgH_2$ signal persists longer where the lamellar structures are absent (**Figure 3**, solid line).

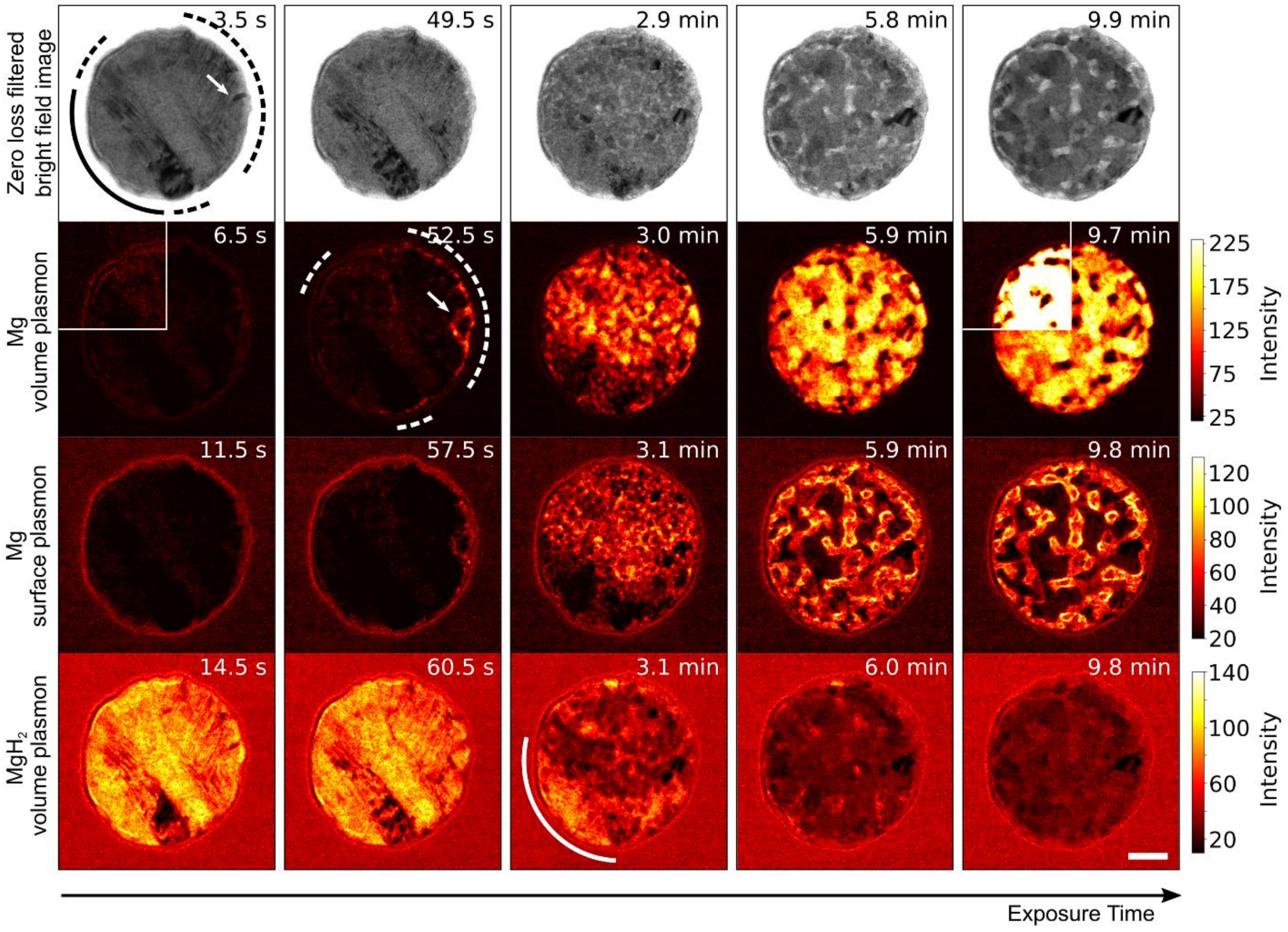

**Figure 3.** ESI series of the *in-situ* dehydrogenation. The time series shows zero loss filtered bright field images (top row), the signal from excitation of the Mg volume plasmon (second

row; imaged with electrons of 10.6 ± 0.5 eV energy loss), of the Mg surface plasmon (third row; 8.0 ± 0.5 eV), and of the $MgH_2$ volume plasmon (bottom row; 15.0 ± 0.5 eV). Intensities are adjusted individually for each signal to achieve the best visibility for all components. Insets in the Mg volume plasmon images shows the signal scaled to the intensity range of the Mg surface plasmon signal. Cumulative exposure time increases from left to right. Arrows and dashed lines indicate fast dehydrogenation, the solid line slow dehydrogenation. Time has been chosen as reaction coordinate as it is more intuitive and the characteristic dose depends on the electron energy[23]. Full image series with electron dose is provided as SI-1 (movie). Scale of 100 nm applies to all images.

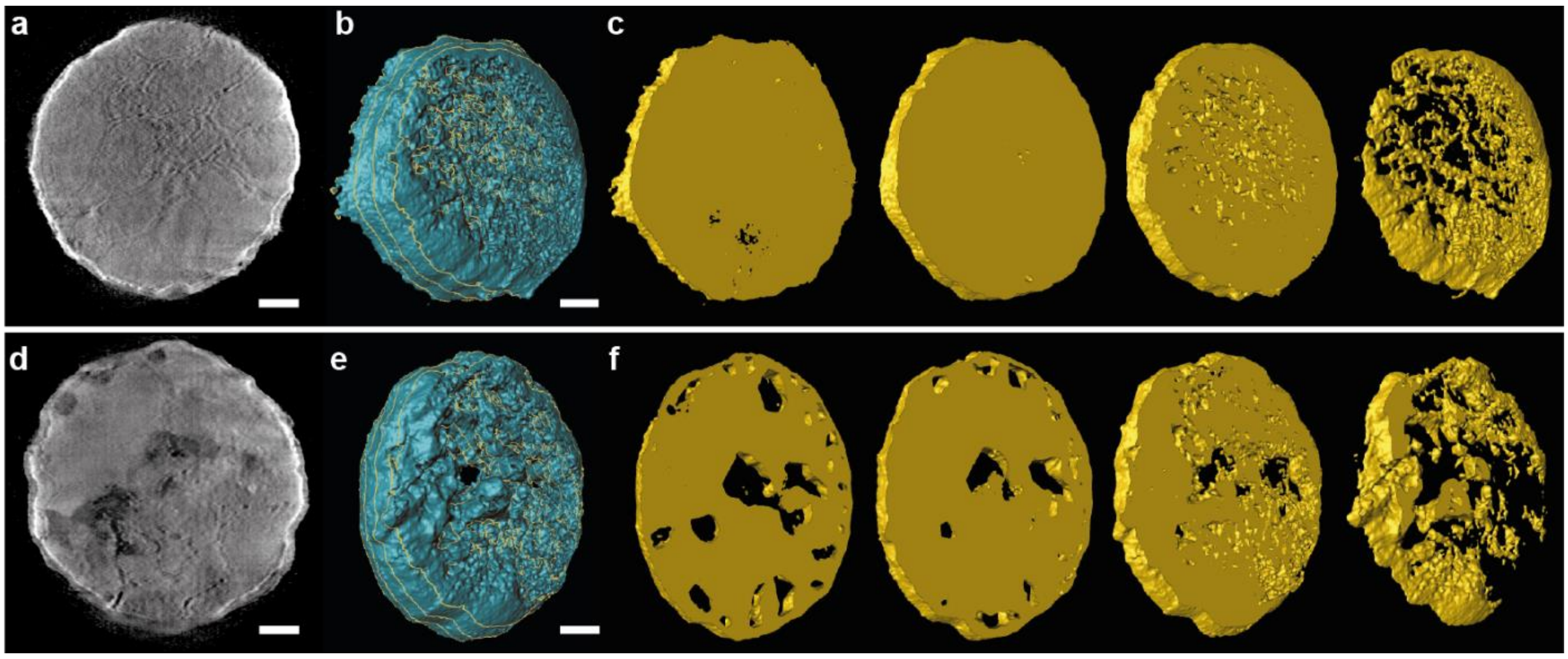

**Figure 4.** Three-dimensional reconstructions of pristine (**a**-**c**) and dehydrogenated (**d**-**f**) nanoparticles from electron tomography. **a, d**, Slice through top of the reconstructions (for **a**: same NP as in Figure 1b). **b, e**, Volume rendering of the NPs. Yellow lines indicate slices (**c**, **f**) to visualize inner voids. Scale of 50 nm applies to all panels.

### 2.3. A Lamellar β-γ-Phase Mix

We used NBD to identify the lamellar defects as a mix of β-$MgH_2$ nanotwins and γ-$MgH_2$ nanolamellas (see **Figure 5**). All NBD patterns show elongated streaks instead of the expected

Bragg spots (**Figure 5d, f** and **S7**, SI). These streaks are perpendicular to the lamellar defects in the corresponding real space image. The degeneration of discrete spots to elongated streaks indicates planar defects or platelet like crystals with limited spatial extension, i.e. large extension in reciprocal space, in the direction of streaking. The power spectra of the lamellar defect structures show continuously enhanced spatial frequencies typically down to 2-3 nm, sometimes as small as 1 nm (e.g. **Figure 5c** with 2.3 nm). This highlights the small and nonuniform thickness of the twin and γ lamellas.

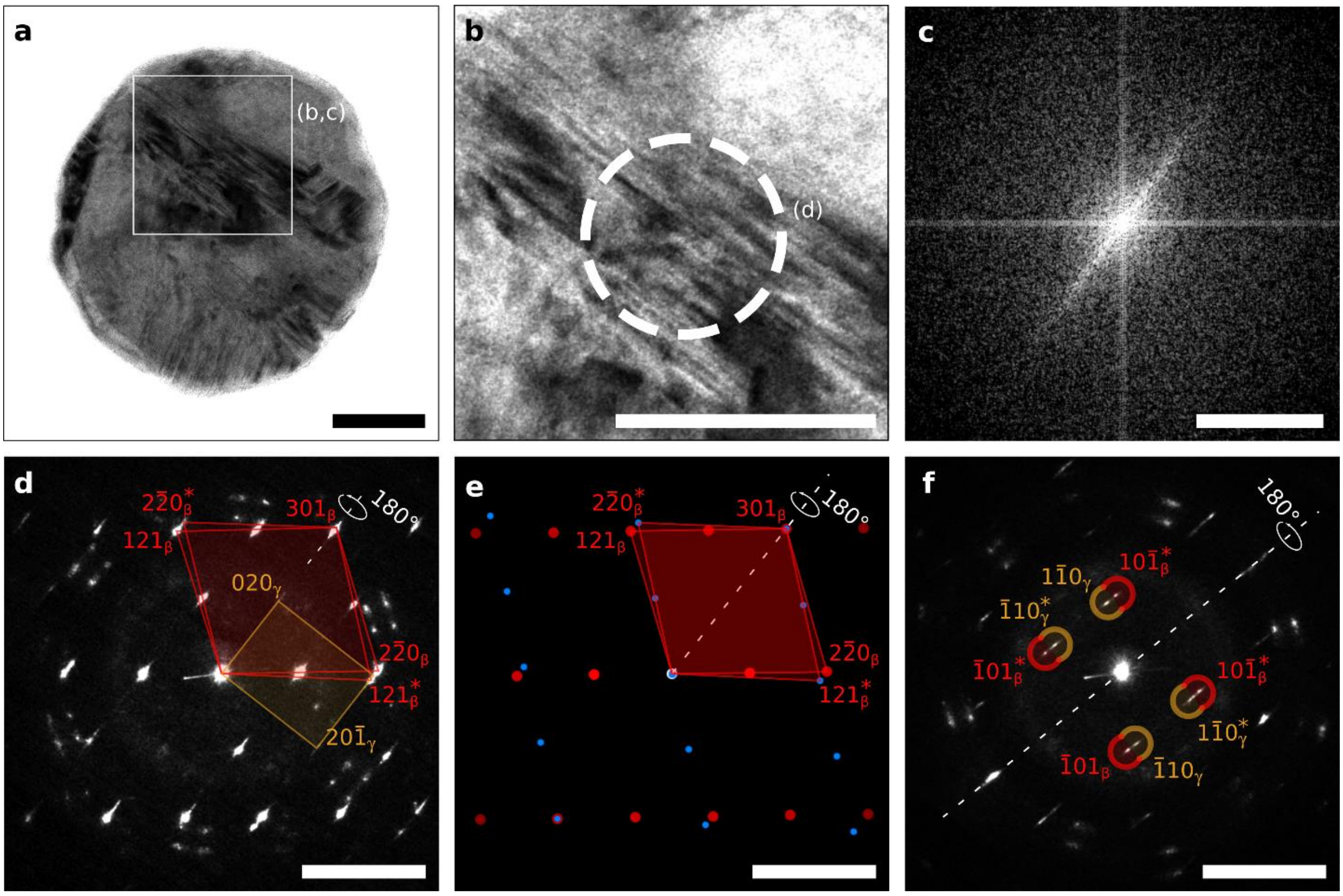

**Figure 5.** Structural analysis of single lamellar defect structures. **a**, Hydrogenated NP with lamellar defect structures. Square indicates the window corresponding to **b**; scale: 100 nm. **b**, Dashed circle indicates the selected volume for the NBD diffraction pattern in **d**; scale: 100 nm. **c**, Power spectrum of **b**; scale 0.5 nm$^{-1}$. **d,** NBD pattern of the lamellar defect structure indexed as a $(301)_\beta$ twin (red) close to the $[11\text{-}3]_\beta$ zone axis. The characteristic 180° rotation symmetry around the $(301)_\beta$ plane normal in direction of streaking is indicated together with γ-$MgH_2$

reflections (yellow) from a zone axis close to $[102]_\gamma$. * denotes reflections originating from 180° rotation symmetry. **e**, Simulated diffraction pattern of a $(301)_\beta$ twin viewed along the $[11\text{-}3]_\beta$ zone axis with matrix (red) and twin (blue) reflections. **j**, NBD pattern from a different hydrogenated NP (bright field TEM image in Figure S8, SI). α and γ reflections are indicated together with the observed 180° rotation symmetry along the direction of streaking. Scale of 5 $nm^{-1}$ applies to all diffraction patterns.

The streaking of the NBD patterns inhibits precise location of the diffraction spots, and thus phase identification. Therefore, we averaged 58 NBD patterns from different lamellar defect structures to form an artificial ring pattern (**Figure 6**). The nanolamellas are mainly composed of rutile β-$MgH_2$ together with its high pressure γ phase with orthorhombic α-$PbO_2$ crystal structure.[33] Due to the orientation relationship between Mg and $MgH_2$[20], the hydrogenated NPs remain textured, which is evident from the overexpression of the $110_\beta$ reflections. In addition, we observe a diffuse ring from nanosized cubic $TiH_2$, together with reflections from an unidentified compound with lattice spacings of 0.160, 0.139, and 0.090 nm. Refer to SI-9 for further interpretation.

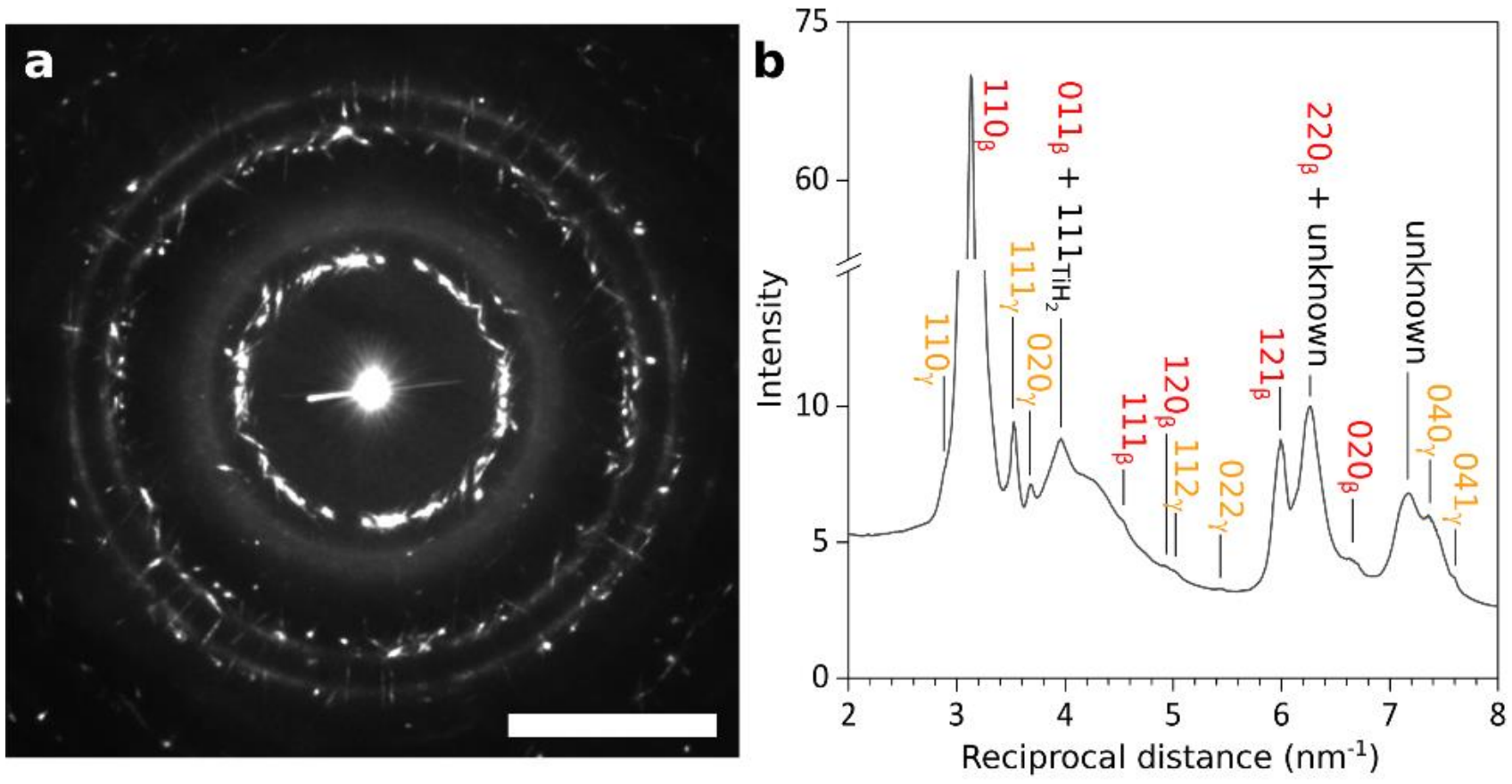

**Figure 6.** Phase composition of the lamellar defect structures from electron diffraction. **a**, Average of 58 nano beam diffraction patterns from lamellar defect structures (cf. Figure 3b).

The intense needle like streak through the zero beam is caused by the opening beam shutter when working in low dose conditions; scale 5 $nm^{-1}$. **b**, Corresponding radial averaged intensity profile with indicated peaks. Refer to SI-10 for the determined lattice parameters.

Knowing the composition, we were able to index individual NBD patterns and identify β-$MgH_2$ $(301)_\beta$ nanotwins as a major structural feature. **Figure 5d** shows the NBD pattern of a $(301)_\beta$ twin close to the $[11\text{-}3]_\beta$ zone axis. It shows the characteristic 180° rotation symmetry around the $301_\beta$ reflection, i.e. the twin plane and direction of streaking, and agrees perfectly with the simulation (**Figure 5e**). In addition, several weak γ-$MgH_2$ reflections are present in the NBD pattern of the $(301)_\beta$ twined NP, e.g. $201_\gamma$ and $020_\gamma$ (further examples in **Figure S7**, SI). The accordingly streaked γ-$MgH_2$ and β-$MgH_2$ reflections indicate parallel β twin- and γ nanolamellas, which is also suggested by the lamellar contrasts in the real space images. This does do not only indicate an orientation relationship between the low-pressure β and the high-pressure γ phase, but also a key role of the β-$MgH_2$ twins for the formation of γ-$MgH_2$, as such ordered structures do not form randomly.

## 3. Discussion

### 3.1. γ-$MgH_2$ Nucleates at β-Twins

The structural features of the NPs in the pristine, hydrogenated and dehydrogenated states are summarized in **Figure 7**. We have identified the lamellar $MgH_2$ defect structures as a mix of $(301)_\beta$ β-$MgH_2$ twins and parallel γ-$MgH_2$ nanolamellas. So far, only the $(101)_\beta$ twins were observed after the hydrogenation of freestanding Mg thin films[30] and deformation twins in ball milled $MgH_2$ powders[21]. However, both the (301) and (101) planes have been reported as stable twin planes for rutile crystals.[34] At this point, we could unambiguously identify $(301)_\beta$ twins in some $MgH_2$ NPs, however, cannot exclude the presence of $(101)_\beta$ twins in others.

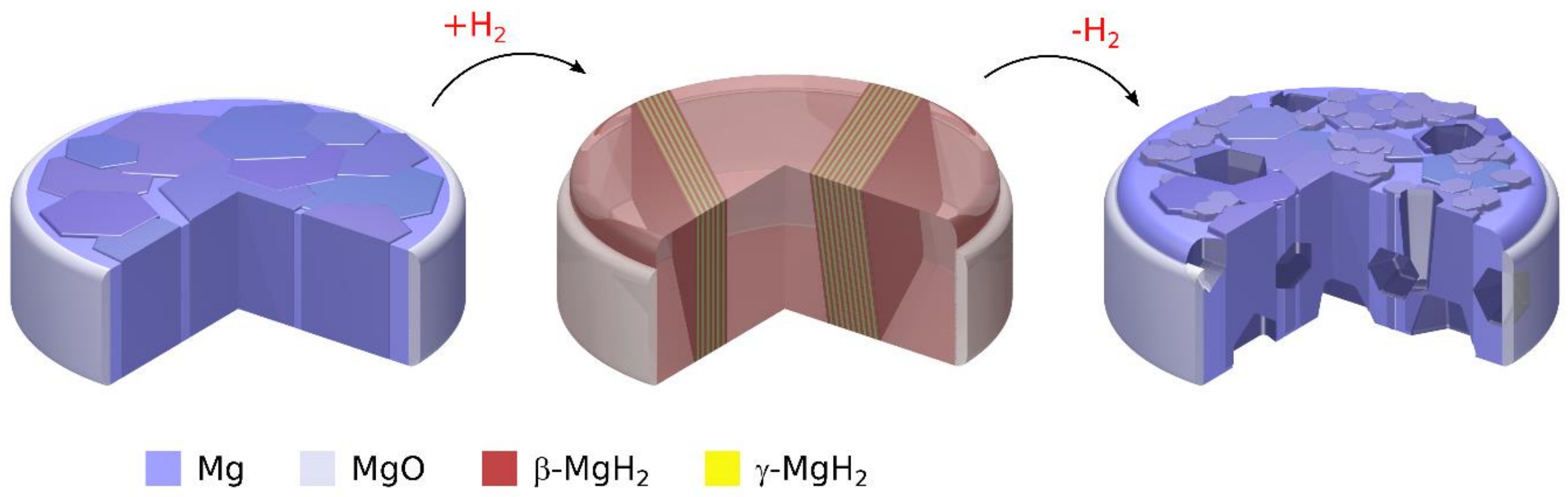

**Figure 7.** Structural characteristics of the NPs in the pristine (**left**), hydrogenated (**middle**) and dehydrogenated (**right**) state. The NPs are covered by a MgO belt (gray). **a**, The pristine polycrystalline NPs show typical Mg grain sizes between 40-90 nm. They are textured with the $[0001]_{Mg}$ direction normal to the substrate (not shown). **b**, The hydrogenated NP consist of large β-$MgH_2$ crystals, which are penetrated by a mix of parallel $(301)_\beta$ nanotwins and γ-$MgH_2$ nanolamellas. The particles remain textured due to the orientation relationship between Mg and $MgH_2$. The MgO belt restricts the volume expansion during hydrogenation to 1D. **c**, The dehydrogenated NPs form voids to accommodate the volume shrinkage during dehydrogenation. These voids form primarily at the bottom. Some transverse the NP completely in normal direction. The preferential void formation at the MgO belt is not observed in smaller NPs (cf. Figures 4f and S6, SI). $Si_3N_4$ substrate and Ti and Pd layers are omitted for clarity.

$MgH_2$ does not only share the same rutile crystal structure with $TiO_2$, but also the same high-pressure configuration with α-$PbO_2$ crystal structure. Meng et al.[35] determined an orientation relationship between rutile $TiO_2$ and its α-$PbO_2$ configuration, where the formation of the high pressure phase is associated with $(101)_{rutile}$ twins, which contain a basic unit of the α-$PbO_2$ crystal structure.[36] This leads to the nucleation of thin high-pressure phase lamellas at and parallel to the $(101)_{rutile}$ twins under high geological compressive stress above 4 GPa. We could not confirm this orientation relationship for the $MgH_2$ NPs due to intense streaking of the diffraction spots, dynamic diffraction and fast $MgH_2$ decomposition. However, due to parallel $\beta$ and $\gamma$ lamellas, a similar mechanism is plausible for the formation of $\gamma$-$MgH_2$.

We support this assumption by the NBD pattern in **Figure 5f**. This NBD pattern shows four different $101_\beta$ and $110_\gamma$ reflections. Not all these reflections can angle wise belong to the same $\beta$ or $\gamma$ zone axis, i.e. to the same crystals. Instead, they are formed by 180° rotation symmetry around the common streaking direction, which would suggest twinning of the $\gamma$-phase as well. Yet, to our knowledge, no twinned $\alpha$-$PbO_2$ type crystals are reported. Thus, it is more likely that the 180° rotation symmetry of the $\gamma$ phase is pseudo and caused by its nucleation at both, matrix and twin lamellas, which inherit 180° rotation symmetry with each other. Hence, this 180° rotation symmetry is inherent for different $\gamma$ lamellas with an orientation relationship towards either the parent $\beta$-$MgH_2$ matrix or the twin. To confirm this hypothesis, further investigation is required by high resolution TEM under cryogenic conditions.

### 3.2. Mg-Diffusion is the Primary Degradation Mechanism of Constrained Mg/$MgH_2$-Nanostructures

We identified Mg diffusion as mechanism of void formation during dehydrogenation. By the formation of voids, the NPs adapt volume shrinkage during dehydrogenation as contraction is inhibited by the MgO belt. Similarly, voids have been observed after the dehydrogenation of $Mg_2NiH_4$ NPs with MgO shell[37] and $MgH_2$ thin films[38], where they primarily form at the interface with the substrate. For thin films, plastic deformation has been suggested as mechanism for structural relaxation after dehydrogenation.[39] However, our results reveal a diffusion governed process. The voids presumably accumulate at the substrate and MgO belt to reduce interfacial energy with the corresponding material. Notably, significant Mg diffusion happens here at room temperature, as diffusion is enhanced by defects formed during dehydrogenation.[23]

### 3.3. γ-$MgH_2$ Improves Dehydrogenation

Dehydrogenation is fastest at grain boundaries, followed by the lamellar structures with twins and γ-$MgH_2$. In both cases, fast dehydrogenation might be assigned to the role of grain, twin and phase boundaries as hydrogen diffusion paths and nucleation sites for Mg.[2,40] Yet, we expect a strong influence from γ-$MgH_2$ itself. One may argue that radiolytic dehydrogenation in the TEM is of small technical relevance and does not entirely reflect the nature of thermal dehydrogenation.[21] However, metastable γ-$MgH_2$ offers faster hydrogen diffusion[41], decomposes at lower temperatures[16–18,37] and destabilizes adjacent β-$MgH_2$ and catalyses its decomposition[16,42]. Thus, stabilizing γ-$MgH_2$ is considered a fundamental aspect of a future strategy to develop Mg based solid state hydrogen storage materials with optimized dehydrogenation.[3,18,42] γ-$MgH_2$ can be formed by severe plastic deformation of β-$MgH_2$.[17,18,42,43] So far, it could only be directly generated during hydrogenation in NPs from electrochemical synthesis but not retained during cycling.[16] Here, we report a novel generation mechanism of a nanostructured β/γ-$MgH_2$ phase mix during hydrogenation, which is based on the inherent volume expansion during hydrogenation.

### 3.4. The Importance of Particle Design

In a recent *in-situ* TEM investigation on thin Mg samples[44], Hamm *et al*. emphasised the role of stress on the structural evolution during hydrogenation. Whereas the evolving $MgH_2$ is under compressive stress, local tensile stress around its growing front leads to plastic deformation of adjacent Mg, resulting in nano grains with small angle grain boundaries. Even though their sample displayed the same texture as our NPs, we cannot confirm their observation. Instead, we observe twins, which cannot be pre-formed in the Mg phase. Therefore, it is the plasticity of the hydride itself that determines the structure of more confined NPs in the hydrogenated state. We consider the plastic deformation via twinning an interplay between the anisotropy of the textured NPs and their complex stress state upon hydrogenation.

In thin films, the substrate or a capping layer mechanically constrain the storage layer in-plane. A similar constrain can be expected from the MgO belt in the present NPs. However, neither Hamm et al. nor any others reported γ-$MgH_2$ or excessive twinning in hydrogenated Mg thin films. Thus, the texture and in-plane mechanic constraints cannot cause twinning and γ-$MgH_2$ alone. In fact, such textured NPs expand only by 6 % in-plane and primarily out-of-plane (23 %).[14] However, the MgO belt inhibits free volume expansion in in-plane and out-of-plane direction. Thus, volume expansion during hydrogenation leads to intense compressive stress upon which the NPs deform plastically and solely expand in out-of-plane-direction[45], as the MgO belt remains intact and no significant diameter increase was observed for the hydrogenated NPs (see **Figure 3**). The plastic deformation is accommodated by twinning in the present NP design and reduces the compressive stress until yield strength is reached. The yield strength of $MgH_2$ is estimated as 1.4 GPa in thin films[46] and 2 GPa in comparable NPs[45]. This high remaining compressive stress results in the formation of the high pressure γ phase.[47] The γ phase is metastable at normal pressure and thus persists within the NPs even after further relieve from compressive stress.[48] As excessive $(301)_\beta$ twinning appears to be a unique feature of the investigated NPs, we account their role as nucleation sites crucial for the formation of the γ-$MgH_2$, rather than their higher yield strength compared to thin films. Weather this formation mechanism also applies to the formation of γ-$MgH_2$ by severe plastic deformation has to be investigated.

## 4. Conclusion

We have demonstrated that the inherent volume expansion during hydrogenation can be exploited to create nanostructures in confined $MgH_2$ particles by twinning and to form metastable γ-$MgH_2$. Large compressive stress arises during hydrogenation and leads to plastic deformation, which is governed by $(301)_\beta$ twinning. These twins are proposedly nucleation sites for $\gamma$-$MgH_2$ nanolamellas. The twins and $\gamma$ lamellas have a typical nonuniform minimal thickness of 2-3 nm. We assume that twinning as dominant deformation mechanisms is a result of the anisotropy of the textured NPs in combination with the in- and out-of-plane mechanical constraints from the MgO belt, which restrict the NPs to 1D expansion in out-of-plane direction during hydrogenation. Texture and anisotropy of the NPs are maintained in the hydrogenated state due to the orientation relationship between Mg and $\beta$-$MgH_2$. Based on our observations and earlier studies, we expect twinning and $\gamma$-$MgH_2$ to improve dehydrogenation on a kinetic, and a kinetic and thermodynamic level, respectively. For technical applications reformation of these defect structures upon each rehydrogenation, i.e. a cycle stable design, must be established. This is possible, as the twins and $\gamma$ lamellas form intrinsically during hydrogenation, but structural degradation by void formation of the NPs must be avoided. A possible approach would be to detach the constraining belt from the NP and enable free contraction, e.g. by incorporating the NPs in a rigid framework. The different surface energies might be used to suitably orient the anisotropic particles.[29] A different approach, suggested by our tomography results, is to control void formation by interfacial effects, as they strongly influence the morphology of voids. Furthermore, the influence of alloying on $\gamma$-$MgH_2$ quantity and stability has to be explored.

## 5. Experimental Section/Methods

*Fabrication of Mg/$MgH_2$ nanoparticles:* We manufactured the samples using electron-beam lithography (EBL). First, we spin-coated (5 s at 3000 r.p.m. and 30 s at 8000 r.p.m., respectively) a double-layer PMMA resist (250k-2.5 % and 950k-1.5 %, Allresist) on a TEM grid (15 nm silicon nitride membrane on a Si frame with 200 µm thick and 3 mm in diameter, PLANO), followed by backing (5 min at 150 °C) on a hot plate. We performed EBL (Raith eLine Plus) with 20 kV acceleration voltage and a 20 µm aperture. After development (90 s in MIBK and 60 s in isopropanol), we applied an oxygen plasma treatment (15 s) to remove PMMA residues. Then we successively deposited, Ti (3 nm), Mg (45 nm), Ti (2 nm), and Pd (3 nm) through electron-beam evaporation (PFEIFFER Vacuum, PLS-500). The deposition rates for Mg, Ti, and Pd were 1.0, 0.1, and 0.2 nm/s, respectively. We used acetone as solvent for metal lift-off (3 hours at room temperature). Afterwards, we coated the backside of the TEM membrane (Vacuum Coater, Leica EM ACE200) with 5 nm carbon to avoid charging during the TEM measurements. We transferred and stored the samples in dry environment until hydrogenation by exposing the sample to hydrogen (5.0 vol.%) with nitrogen carrier gas at room temperature in a custom-made gas chamber.

*Transmission electron microscopy:* For the TEM investigations, we used a monochromated and $C_s$-corrected Zeiss Libra 200® (Carl Zeiss Microscopy GmbH) with stable Omega filter.[49] We transferred the samples in air. We performed electron spectroscopic imaging, electron energy loss spectroscopy and nanobeam diffraction at 60 kV acceleration voltage. At 60 kV, the dose rate was about 1,200 e nm$^{-2}$ s$^{-1}$.

*Electron spectroscopic imaging:* For ESI, we used an angle limiting objective aperture and an energy selective slit of 1 eV width, centred on zero energy loss. We subsequently and repetitively increased the primary energy by 10.6 eV, 8 eV, 15 eV and 0 eV to image the corresponding energy loss with exposure times of 3.0 s, 5.0 s, 3.0 s and 0.50 s, respectively. Additionally, we acquired an initial and final zero loss image with 2.0 s exposure. For one cycle of ESI images, the electron dose accumulated to 13,800 e nm$^{-2}$. We denoised all images with a median filter of $r$=3 pixel and an absolute threshold of 20, followed by factor 4 binning and smoothed the displayed ESI images further with a Gaussian blur of $r$=1.5 pixel.

*Nanobeam diffraction:* For NBD[26], we used a parallel probe of ca. 80 nm diameter. Previoulsy, we imaged the NP with a zero-loss filtered bright field image (0.50 s exposure) to locate the defect structures and centred the probe by micro dose focusing, where the beam is tilted around the NP and further illumination avoided. We acquired eight zero loss filtered patterns per defect structure with an exposure time of 500 ms. We denoised the NBD patterns with a median filter of *r*=3 pixels and an absolute threshold of 20. We subsequently aligned them along the zero beams, averaged them, and corrected them for residual astigmatism all with a custom script. Finally, we aligned the rotation of the diffraction patterns and the real space images. In order to form the artificial ring patten, we averaged all acquired patterns after aligning them along the zero beams. We used JEMS SAAS for the simulation of the diffraction patterns and the twin orientation relationships [010]//[010],(101)//(10-1) and [010]//[010],(301)//(-301), which we derived from Danaie *et al*.[21] and Lee *et al*.[34]

*Electron energy loss spectroscopy:* For EELS, we used an angle limiting objective aperture, obtaining an energy resolution between 100 eV and 145 eV with monochromator. We used a custom script including the HyperSpy toolbox[50] to process the spectrum images. For each time step, we averaged 10 spectra with an exposure time of 200 ms each, of which was each shifted on the camera by increasing the primary energy by 0.5 eV to reduce static noise. We exposed the NP for 18 s between each of the EELS series, which results in time steps of 20 s. We denoised the images with a median filter of r=3 pixel, an absolute threshold of 20, and a relative threshold of 3.7. We found the latter to conserve energy resolution. We used the -Lucy method[51] with 6 iterations and a $Si_3N_4$ spectrum (cf. SI-5) as kernel to remove its contribution to the low loss EEL spectra.

*Electron tomography:* We conducted electron tomography at 200 kV acceleration voltage and 5 days after dehydrogenation with a Saxton scheme[52], ± 64° tilt, and without objective aperture to reduce diffraction contrast. We aligned the images of the tilt series by their centre of mass with a custom script and reconstructed the volume using the ASTRA toolbox.[53] We ascribe the large number of 10k iterations of the applied SIRT algorithm to account for the missing wedge in connection to the low contrast to the $Si_3N_4$ membrane and residual diffraction contrast. Segmentations of the reconstructed volumes were superior to using weighted back projection and segmentations from Discrete Algebraic Reconstruction Technique (data not shown). The reconstructed volume was rendered and segmented in Amira (ThermoFisher Scientific, USA).

**Supporting Information**

Supporting Information is available from the Wiley Online Library or from the author.

**Conflict of Interest**

The authors declare no conflict of interests.

**Author Contributions**

J.A.K., X.D., F.N., N.L., and M.P. conceived the project. J.A.K. performed the TEM investigations and data analysis under supervision of M.P. and R.R.S. X.D. fabricated the samples. J.A.K., X.D., and M.P. drafted the manuscript. All authors discussed the results and revised the manuscript.

**Acknowledgements**

This project has received funding by the Ministry of Science, Research and the Arts Baden-Württemberg, through the HEiKA materials research centre FunTECH-3D (MWK, 33-753-30-20/3/3). J.A.K. and R.R.S. have been funded by the Deutsche Forschungsgemeinschaft (DFG, German Research Foundation) under Germany's Excellence Strategy via the Excellence Cluster 3D Matter Made to Order (EXC-2082/1 – 390761711). X.D., F.N., and N.L. acknowledge support by the European Research Council (ERC Dynamic Nano) grant. N.L. acknowledges support by the Max Planck Society (Max Planck Fellow). The authors acknowledge the data storage service SDS@hd supported by the Ministry of Science, Research and the Arts Baden-Württemberg and the German Research Foundation (DFG) through grant INST 35/1314-1 FUGG. N.L.

**Table of Content**

We demonstrate the intrinsic formation of metastable γ-$MgH_2$ during hydrogenation of suitably designed Mg nanoparticles. We propose a formation mechanism and identify the primary degradation mechanism of the nanoparticles based on transmission electron microscopy. The combined knowledge of both mechanisms is expected to enable cycle stable, Mg based materials for dynamic plasmonics and hydrogen storage, as γ-$MgH_2$ enhances dehydrogenation kinetics and thermodynamics.

J.A.K. Author 1, X.D. Corresponding Author 1*, F.N. Author 3, R.R.S. Author 4, N.L. Corresponding Author 2*, M.P. Corresponding Author 3*

**Stabilizing γ-$MgH_2$ at Nanotwins in Mechanically Constrained Nanoparticles**

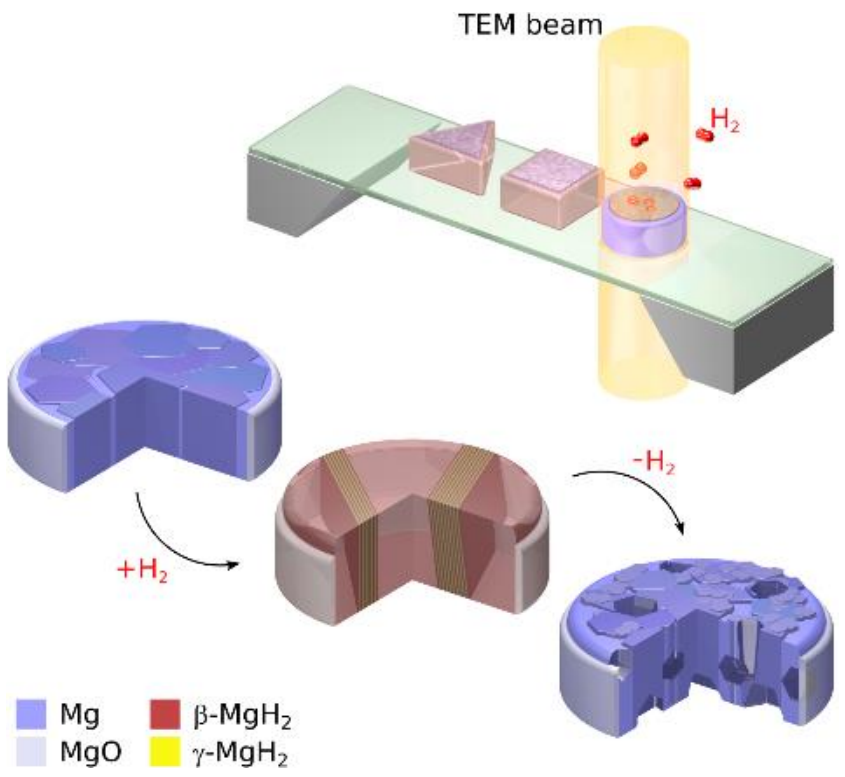

ToC figure, size: 55 mm broad × 50 mm high